\documentclass[aps,prd,twocolumn,showpacs,floatfix,superscriptaddress,
preprintnumbers,nofootinbib]{revtex4}

\usepackage{epsfig,amssymb}

\usepackage[dvips]{color}
\definecolor{Black}{named}{Black}
\definecolor{Red}{named}{Red}
\definecolor{Blue}{named}{Blue}

\def\lsim{\raise0.3ex\hbox{$\;<$\kern-0.75em\raise-1.1ex\hbox{$\sim\;$}}}
\def\gsim{\raise0.3ex\hbox{$\;>$\kern-0.75em\raise-1.1ex\hbox{$\sim\;$}}}

\def\theta{\vartheta}
\def\tdiff{t_{\rm diff}}

\def\lsyn{l_{\rm syn}}

\newcommand{\be}{\begin{equation}}
\newcommand{\ee}{\end{equation}}
\newcommand{\ba}{\begin{eqnarray}}
\newcommand{\ea}{\end{eqnarray}}

\begin{document}

\preprint{IFIC/07-33}

\title{High energy neutrino yields from astrophysical sources II:
Magnetized sources}

\author{M.~Kachelrie\ss}
\affiliation{Institutt for fysikk, NTNU, N--7491 Trondheim, Norway}

\author{S.~Ostapchenko}
\affiliation{Institut f\"ur Experimentelle Kernphysik,\\
Universit\"at Karlsruhe, 76021 Karlsruhe, Germany}
\affiliation{D.~V.~Skobeltsyn Institute of Nuclear Physics,\\
 Moscow State University, 119992 Moscow, Russia}

\author{R.~Tom\`as}
\affiliation{AHEP Group, Institut de F\'{\i}sica Corpuscular -
C.S.I.C./Universitat de Val\`encia,
\\ Edifici Instituts d'Investigaci\'o, Apt. 22085, 
E--46071 Val\`encia, Spain}

\date{August 22, 2007}

\begin{abstract}
We calculate the yield of high energy neutrinos produced in astrophysical 
sources for arbitrary interaction depths $\tau_0$ and magnetic field
strengths $B$. We take into account energy loss processes like synchrotron
radiation and diffusion of charged particles in turbulent magnetic fields 
as well as the scattering of secondaries on background
photons and the direct production of charm neutrinos. Meson-photon
interactions are simulated with an extended version of the SOPHIA
model. Diffusion leads to an increased path-length before protons
leave the source of size $R_s$ and therefore magnetized sources lose
their transparency below the energy 
$E\sim 10^{18}{\rm eV} \: (R_s/{\rm pc}) \: (B/{\rm mG})
\tau_0^{1/\alpha}$, with $\alpha=1/3$ and 1 for Kolmogorov and Bohm
diffusion, respectively. Moreover, the neutrino flux is suppressed
above the energy where synchrotron energy losses become important for
charged particles. As a consequence, the energy spectrum and the
flavor composition of neutrinos are strongly modified both at low and
high energies even for sources with $\tau_0\lsim 1$.  
\end{abstract}


\maketitle

\section{Introduction}

High energy neutrinos from astrophysical sources are the decay products of
secondary mesons produced by scattered  high energy protons on background
protons or photons.  A classic example are the so-called cosmogenic or GZK
neutrinos produced in scatterings of extragalactic ultra-high
energy cosmic rays on cosmic microwave photons during 
propagation~\cite{GZKnu}. Since the sources of ultra-high energy 
cosmic rays unavoidably contain also matter and photons, all cosmic
ray sources like active galactic nuclei (AGN) or gamma-ray bursts should  
produce to some extent also high energy neutrinos~\cite{agn,agnII,agnI,grb}. 
However, the elusive nature of their interactions has prevented so far
the detection of neutrinos from other sources than the Sun and
SN1987A, despite of intensive experimental efforts. 

Theoretical limits or predictions for neutrino fluxes are therefore 
important as a guideline for experimentalists.
Two different kinds of bounds on
high energy neutrino fluxes exist: The cascade or EGRET limit uses bounds on
the diffuse MeV-GeV photon background to limit the energy transferred to
electromagnetically interacting particles that are produced unavoidably
together with neutrinos~\cite{casc}. The cosmic ray upper
bounds of, e.g., Refs.~\cite{WB,MPR} use the observed ultra-high
energy cosmic ray flux to restrict possible neutrino fluxes. 
One of the several underlying assumptions of the latter limit is that
all neutrino sources are transparent to hadronic interactions and thus
at least neutrons can escape from the source region without interactions.  

In a previous work, we calculated the flux of high energy
neutrinos produced as secondaries in astrophysical sources with
arbitrary interaction depth but negligible magnetic fields~\cite{I}. 
The present work extends Ref.~\cite{I} in two respects:
First, multiple interactions that become relevant in a thick source
require the modeling not only of photo-nucleon ($N+\gamma\to X$), but
also of photo-meson ($\pi+\gamma\to X$, $K+\gamma\to X$) interactions. 
The latter are treated now within a self-consistent extension of the 
SOPHIA model~\cite{sophia}. 
Second, and more importantly, we have added the effects
of possible magnetic fields around the source region. This includes
diffusion of charged particles in turbulent magnetic fields and energy
loss processes like synchrotron radiation. As a result, the shape of
the neutrino spectra deviates strongly from  the injection spectrum of
protons even for transparent sources, $\tau_0\lsim 1$, both at the low
and the high energy end of the produced neutrino flux. Moreover, the
relative importance of the various channels contributing to the
neutrino yields changes strongly, leading thereby to large variations
in the neutrino flavor composition as function of the energy. 
Specific examples are the cases where the neutrino spectrum is
dominated by kaon decays~\cite{Ando:2005xi,Asano:2006zz} or influenced
by muon damping~\cite{Kashti:2005qa}. 

This work is structured as follows: In Sec.~II, we discuss how we
simulate the interactions and the propagation   of charged
particles. In particular, we describe our extension of the SOPHIA
model, the continuous energy loss processes considered and how we 
account for diffusion. In Sec.~III, we discuss the influence of
magnetic fields on the escaping proton flux. Armed with our
understanding of the proton fluxes, we continue in Sec.~IV to examine
the resulting neutrino fluxes and in Sec.~V their flavor composition.
Finally, we summarize our results in  Sec.~VI.

The application of our simulation to concrete astrophysical models
and a comparison with earlier studies of neutrino fluxes from
AGN cores~\cite{agnII} and jets~\cite{agnI}, clusters of
galaxy~\cite{cluster}, or hidden sources~\cite{hidden}
will be performed in paper III of this series~\cite{III}.

\section{Simulation of particle interactions and propagation}

\subsection{Simulation of photon-hadron interactions}

\subsubsection{Photo-nucleon interaction in the SOPHIA model}

The SOPHIA model~\cite{sophia} was designed to treat gamma-nucleon
interactions in the broad energy range from the production threshold of
mesons up to TeV center of mass (c.m.) energies $\sqrt{s}$. The underlying
physics changes substantially over this energy range. At low energies,
typically below $\sqrt{s}\sim 1\div2$~GeV, the interaction process is
dominated by the excitation of resonant states of the incident
nucleon and by the so-called direct meson production process. On the
other hand, at higher energies the process is dominated by the $t$-channel
exchange of composite states, Reggeons and Pomerons, leading
to multiple production of secondary hadrons.

The cross section for $s$-channel resonance production is described
by the Breit-Wigner formula and results in a strong energy dependence
with pronounced resonance peaks. All the basic parameters like resonance
mass, width, branching ratios as well as angular distributions for
resonance decays are taken from experimental data; when no information
on the angular distributions was available, an isotropic decay was assumed.

Direct meson production corresponds to a $t$-channel exchange of
a virtual meson state, which implies a strong coupling with the nucleon
and an electromagnetic (e/m) one with the photon. As a consequence,
neutral meson exchanges are strongly suppressed and one deals with
charge exchange processes; dominant contributions in case of $\gamma p$
collisions are $\gamma p\rightarrow\pi^{+}n$, $\gamma
p\rightarrow\pi^{-}\Delta^{++}$, 
the $\gamma n$ case follows from isospin symmetry. The contribution
of heavier meson and nucleon exchanges are strongly suppressed as
$1/m_{t}^{2}$, 
$m_{t}$ being the mass of the exchanged state, and can be neglected.
Still, with increasing $s$ additional contributions come from other
$t$-channel resonances lying on the Regge trajectory with pion quantum
numbers, so that the corresponding contribution to the cross section
behaves as $s^{2(\alpha_{\pi}(0)-1)}$, where $\alpha_{\pi}(0)\simeq0$
is the intercept of the pion Regge trajectory. Therefore, in the SOPHIA
model the contribution of the direct process to the $\gamma N$ cross
section is described by a smooth function, which vanishes near the hadron
production threshold and behaves like $s^{-2}$ in the high energy
limit. The corresponding normalization was tuned to reproduce
experimental data on binary production processes. The angular distribution
for the direct process is strongly forward peaked and was parameterized
as
\begin{equation}  \label{dsig/dt-dir}
 \frac{d\sigma_{\gamma N}^{{\rm dir}}}{dt}\sim e^{b_{\rm dir}t} \,, 
\end{equation}
where $t$ is the momentum transfer squared in the process. The value
of the slope $b_{{\rm dir}}$ was taken from a fit to experimental
data \cite{donn78}.

At higher energies the interactions are dominated by Reggeon and Pomeron
exchanges between the nucleon and the hadronic state of the photon,
which result in multiple production of secondary hadrons. Both Reggeons
and Pomerons are composite exchanged states which emerge in the so-called
$1/N_{{\rm c}}$, $1/N_{{\rm f}}$ QCD expansions \cite{hooft},
where Reggeons correspond to a set of diagrams characterized by a
planar topology, whereas Pomerons are described by non-planar (cylinder)
diagrams. Qualitatively, one can see a typical Reggeon-dominated interaction
as follows. The incoming photon converts into a vector meson state,
consisting of a {}``dressed'' color-connected quark-antiquark pair.
Next, the antiquark annihilates with a quark from the nucleon, such
that both the spectator quark from the vector meson state and the
spectator diquark from the nucleon are connected by a color field. While
these spectator partons move apart with their initial velocities,
the color field is stretched into a tube (color string) and finally
disrupts, resulting thereby in the production of secondary hadrons,
which is described by the string fragmentation model~\cite{lund}. 
On the other hand, Pomeron exchange does not lead to a
transfer of quantum numbers between the interacting states. Instead,
it induces a color exchange between the valence constituents of the
vector meson and the nucleon, such that color strings are formed
correspondingly 
between the quark from the meson and the diquark from the nucleon,
and between the meson antiquark and the nucleon quark. Thus, Pomeron
exchange results in the production and hadronization of two color
neutral strings, compared to just one for the Reggeon exchange. Apart
from the string fragmentation procedure, one has to specify here the
energy-momentum partition between valence constituents of the interacting
hadronic states. They are defined by the intercepts of secondary
Reggeon trajectories \cite{cap} and the distribution of light
cone momentum fraction of valence quarks ($x$) and diquarks ($1-x$)
in the nucleon is approximated as
\begin{equation} \label{f-q/p}
 f_{q/N}(x)\sim\frac{1}{\sqrt{x}}(1-x)^{1.5} \,,
\end{equation}
while for the valence quarks ($x$) and antiquarks ($1-x$) in mesons
\begin{equation} \label{f-q/pi}
f_{q/\pi(\rho,\omega)}(x)\sim\frac{1}{\sqrt{x(1-x)}} 
\end{equation}
was used.

Multiple Pomeron exchanges can be neglected still  with reasonable
accuracy in the SOPHIA energy range. However, the model takes into
account diffraction production mechanism, which is treated phenomenologically
as a constant share of the Pomeron and Reggeon exchange contributions.
The angular distribution for diffraction interactions is strongly
forward peaked and is parameterized similarly to (\ref{dsig/dt-dir}),
using an energy dependent slope $b_{{\rm diffr}}$ from \cite{donn78}.

\subsubsection{Generalization for photo-meson interactions}

We have performed a direct generalization of the SOPHIA approach for the
case of gamma-meson ($\pi^{\pm}$, $K^{\pm}$, $K_{0}$, $\bar{K}_{0}$)
interactions. As above in the $\gamma N$ case, the dominant mechanisms
are resonance production and binary processes at low
energies and Reggeon (Pomeron) exchanges at high ones. We have included
the known resonances, with well measured e/m branching width, from
the compilation in the Review of Particle Properties \cite{caso}, the
complete list is given in Table~\ref{res-tab}. All the resonances were
assumed to 
decay isotropically. Concerning the direct meson production, the only
difference compared to the $\gamma N$ case is that the virtual meson-nucleon
coupling is replaced by the meson-meson one, as shown in Fig.~1.%

\begin{figure}
\begin{center}
\includegraphics[width=0.45\textwidth,angle=0,clip]{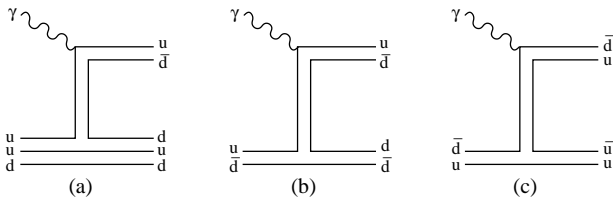}
\end{center}
\caption{\label{direct-pi-fig}
Direct $\pi^{+}$ production in $\gamma p$ (a) and $\gamma\pi^{+}$
(b), (c) interactions.}
\end{figure}

 In the energy range of interest, experimental data on binary processes
in hadron-hadron and gamma-hadron collisions are in a very good agreement
with the additive quark model (AQM) predictions \cite{aqm}. Thus, we
define the meson-hadron coupling as the sum of the contributions
corresponding to the coupling to ``dressed'' constituent quarks.
Correspondingly, for the cross section for the binary process
$\gamma\pi^{\pm}\rightarrow\pi^{\pm}\pi^{0}$ 
we use the SOPHIA parameterization for the reaction $\gamma
p\rightarrow\pi^{+}n$ 
(with properly modified hadron production threshold), based on the
fact that the number of valence (anti-) quarks in pion is the same
as the number of valence $u$-quarks in proton. The same parameterization
is used for $\gamma K^{+}\rightarrow\pi^{+}K_{0}$ (similarly for
other kaons), now being rescaled by a factor $1/2$, corresponding
to having just one light valence \mbox{(anti-)} quark in kaons. Here we neglect
the $t$-channel exchange of kaon states, which is mass squared suppressed
compared to pion exchange, as discussed above.

A remarkable property of Reggeon exchanges is the factorized form
of the corresponding amplitude, which is parameterized as~\cite{collins}
\begin{eqnarray}
 A_{ab}^{\mathbb{R}^{\zeta}}(s,t) &=& 
 8\pi s_{0}\, g_{a}^{\mathbb{R}^{\zeta}}(t)\,
 g_{b}^{\mathbb{R}^{\zeta}}(t)\,
 \eta_{\mathbb{R}^{\zeta}}(t)
 \left(\frac{s}{s_{0}}\right)^{\alpha_{\mathbb{R}^{\zeta}}(t)} 
\label{reg-ampl}
\\
 \eta_{\mathbb{R}^{\zeta}}(t) &=&
 \frac{1+\zeta\,\exp\!\left(-i\pi\,\alpha_{\mathbb{R}^{\zeta}}(t)\right)}
      {\sin\!\left(-i\pi\,\alpha_{\mathbb{R}^{\zeta}}(t)\right)}
 \, , \label{reg-sign}
\end{eqnarray}
where $\alpha_{\mathbb{R}^{\zeta}}(t)$ is the Regge trajectory for
the Reggeon $\mathbb{R}^{\zeta}$, $g_{a}^{\mathbb{R}^{\zeta}}(t)$
the Reggeon-hadron $a$ vertex, and $\zeta=\pm1$ is the Reggeon signature;
$s_{0}\simeq1$~GeV$^{2}$ is the hadronic mass scale. This expression is also
valid for Pomeron exchange, the latter being a particular Reggeon
case, corresponding to the intercept $\alpha_{\mathbb{P}}(0)>1$,
positive signature, and zero transfer of quantum numbers. This allowed
to construct very successful parameterizations for high energy behavior
of hadron-hadron and gamma-proton cross sections \cite{donn92}, which include
only three terms: exchanges of Reggeons with positive ($\mathbb{R}^{+}$)
and negative ($\mathbb{R}^{-}$) signature, and of the Pomeron
($\mathbb{P}$)
\begin{equation}
\sigma_{ab}(s)=C_{ab}^{\mathbb{R}^{+}}\,
s^{\alpha_{\mathbb{R}^{+}}(0)-1}+C_{ab}^{\mathbb{R}^{-}}\,
s^{\alpha_{\mathbb{R}^{-}}(0)-1}+C_{ab}^{\mathbb{P}}\:
s^{\alpha_{\mathbb{P}}(0)-1},\label{sig-reg-pom}
\end{equation}
where the constants $C_{ab}^{X}$ are related to the Reggeon and Pomeron
couplings to hadrons $a$ and $b$, $C_{ab}^{X}\sim g_{a}^{X}(0)\, g_{b}^{X}(0)$,
and the energy exponents of the three terms in (\ref{sig-reg-pom})
are expressed via the intercepts $\alpha_{X}(0)$ of the corresponding
Regge trajectories and do not depend on the types of  interacting
hadrons. Thus, we can obtain the corresponding parameters for
$\gamma\pi$ and $\gamma K$ interactions as
\begin{eqnarray}
C_{\gamma\pi}^{X}& = & C_{\gamma
  p}^{X}\frac{C_{\pi^{+}p}^{X}}{C_{pp}^{X}}\label{C-gamma-pi}\\ 
C_{\gamma K}^{X} & = & C_{\gamma
  p}^{X}\frac{C_{K^{+}p}^{X}}{C_{pp}^{X}},\label{C-gamma-K} 
\end{eqnarray}
for $X=\mathbb{R}^{+},\mathbb{P}$. On the other hand, 
we have $C_{\gamma h}^{\mathbb{R}^{-}}=0$,
as the Reggeons with negative signature do not contribute to $\gamma h$
scattering. The parameters $C_{\gamma p}^{X}$, $C_{hp}^{X}$, $\alpha_{X}(0)$
have been taken from the same cross section fits \cite{caso} which
have been already employed in the SOPHIA model.

\begin{figure}
\includegraphics[width=0.45\textwidth,angle=0,clip]{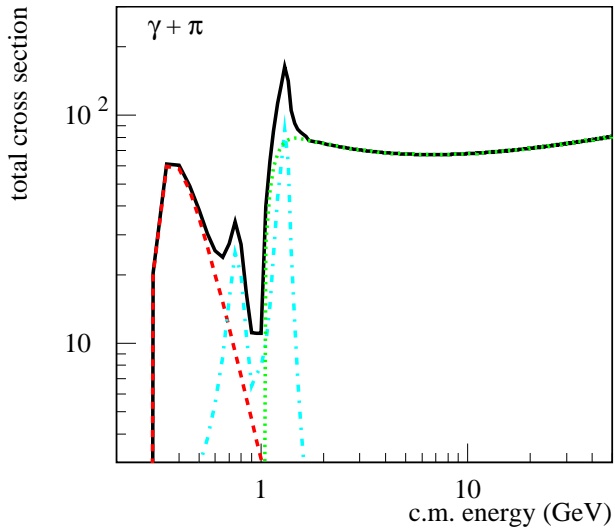}
\caption{\label{gamma-pi-fig}
The $\gamma\pi^{\pm}$  total cross section (solid line)
and the contributions of resonance production (dashed line), of the
direct process (dotted line), and of Reggeon and Pomeron exchanges
(dot-dashed line).}
\end{figure}

\begin{figure}
\includegraphics[width=0.45\textwidth,angle=0,clip]{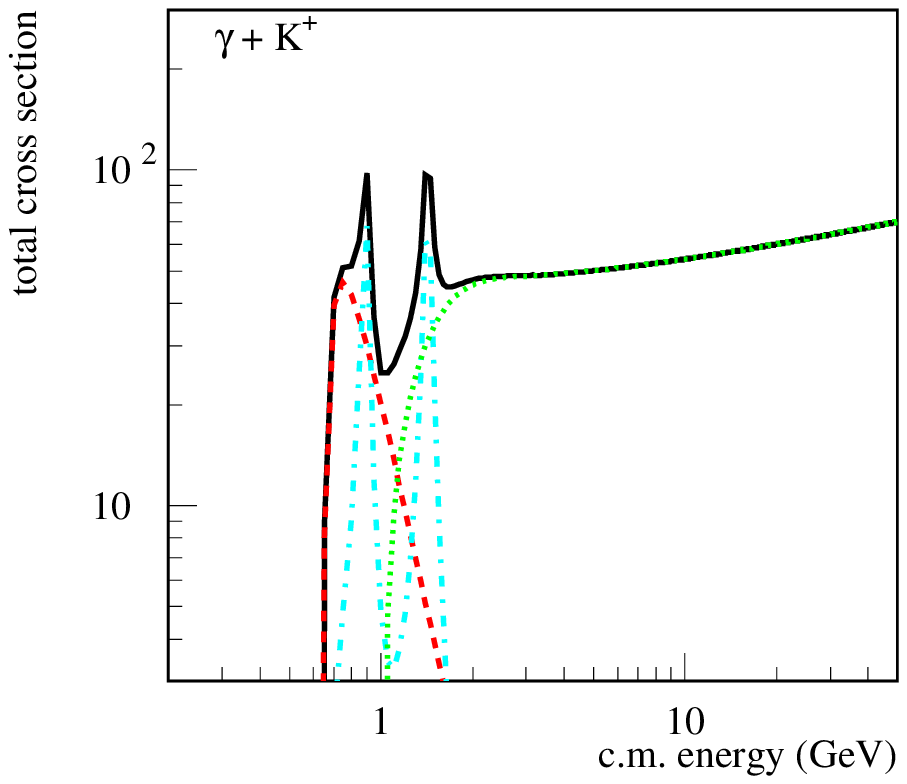}
\caption{ \label{gamma-Kch-fig}
The  $\gamma K^{\pm}$ total cross section and the contributions
of various interaction channels. The abbreviations for the curves
are the same as in Fig.~\ref{gamma-pi-fig}.}
\end{figure}

\begin{figure}
\includegraphics[width=0.45\textwidth,angle=0,clip]{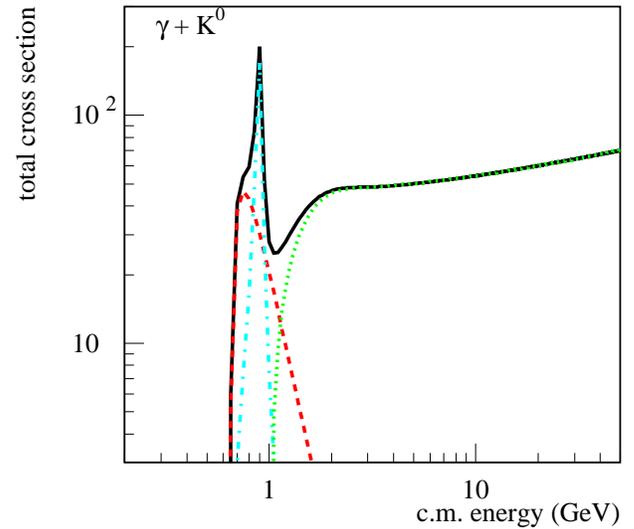}
\caption{\label{gamma-K0-fig}
The $\gamma K_{0}$ total cross section and the contributions
of various interaction channels. The abbreviations for the curves
are the same as in Fig.~\ref{gamma-pi-fig}.}
\end{figure}

The calculated total $\gamma\pi$ and $\gamma K$ cross sections,
as well as partial contributions of various interaction channels,
are plotted in
Figs.~\ref{gamma-pi-fig}--\ref{gamma-K0-fig}.

 In contrast to the $\gamma N$ case \cite{sophia}, production of resonances
does not play a very prominent role in $\gamma\pi$ and $\gamma K$ collisions, 
because of the comparatively high energy thresholds of these processes.
On the other hand, of considerable importance is the direct pion production,
which starts already at the hadron production threshold. It is worth
stressing again that both the direct contribution and the one of Reggeon
and Pomeron exchanges, dominant at high energies, are well-defined
by the proper rescaling from the $\gamma N$ case; the former using
the AQM picture, the latter within the Regge theory framework.

For particle production, we have used as SOPHIA the  Lund JETSET 7.4
string fragmentation procedure \cite{lund}, as well as the distribution
(\ref{f-q/pi}) for the energy-momentum partition between valence
constituents. Also we have kept the same proportion between diffractive
and non-diffractive final states for $\gamma\pi$ and $\gamma K$
interactions as the one for $\gamma p$ case in SOPHIA%
\footnote{One may expect some difference in the diffraction excitation
  probability for pions and kaons compared to protons. The data on
  $pp$, $\pi p$, and $Kp$ interactions indicate that this difference
  is at the 10\% level and therefore inessential for our applications.}.
In summary, the described generalization of the SOPHIA interaction mechanism
for $\gamma\pi$ and $\gamma K$ collisions is well-defined: All the
relevant parameters are fixed either by experimental data or by theoretical
arguments. Finally, we note that in comparison with the simplified
modeling of pion and kaon interactions in Ref.~\cite{I}, the
interaction lengths of pions and kaons are reduced outside the
resonance region by $\sim 35\%$.

\begin{table}[t]\begin{center}
\caption{The meson resonance processes considered and their
  characteristics: mass $M$, width  $\Gamma$, e/m decay width
$b_{\gamma}$, and hadronic decay modes. \label{res-tab}}
\begin{tabular}{|l|c|c|c|c|}
\hline
resonance process  & $M/$GeV  &  $\Gamma/$GeV & $10^3\,b_{\gamma}$ & decay modes \\
\hline
$\gamma \pi^{\pm} \rightarrow \rho^{\pm}$(770)   &  0.768    & 0.150   &  0.45 &
 $\pi\pi$ (100\%)  \\
$\gamma \pi^{\pm} \rightarrow b_1^{\pm}$(1235)   &  1.230    & 0.142   &  1.6 & 
$\omega\pi$ (100\%)  \\
$\gamma \pi^{\pm} \rightarrow a_2^{\pm}$(1320)   &  1.318    & 0.107   &  2.68 &
 $\eta\pi$ (14.5\%)  \\
&&&&  $\rho\pi$ (70.1\%)  \\&&&&  $\omega\pi$ (10.6\%)  \\&&&&  $K\bar K$ (4.8\%
)  \\ 
$\gamma K^{\pm} \rightarrow K^{*(\pm)}$(892)   &  0.892    & 0.0508   &  0.99 & 
$K\pi$ (100\%)  \\
$\gamma K^{\pm} \rightarrow K_2^{*(\pm)}$(1430)   &  1.426    & 0.0985   &  2.4 
& $K\pi$ (50\%)  
\\&&&&  $K^{*}(892)\,\pi$  
\\&&&&  (38\%)  
\\&&&&  $K\rho$ (9\%)  \\&&&&  $K\omega$ (3\%)  
\\ 
$\gamma K^{0} \rightarrow K^{*(0)}$(892)   &  0.896    & 0.0507   &  2.3 & $K\pi
$ (100\%)  \\
\hline
\end{tabular}\end{center}
\end{table}

\subsection{Interactions in the continuous energy loss approximation}

The following interactions are treated in the continuous energy loss
approximation. 

\begin{enumerate}
\item
Synchrotron radiation:
Processes in external magnetic fields are characterized by the
parameters $\chi=eF_{\mu\nu}p^\mu p^\nu/m^3=(p_\perp/m) (B/B_{\rm cr})$ 
and $B_{\rm cr}=m^2/e$.  
The intensity of synchrotron radiation can be approximated for
$B\ll B_{\rm cr}$ by~\cite{ST}
\ba 
 \frac{W}{W_{\rm cl}} &=& 1-\frac{55\sqrt{3}}{24}\xi+\frac{64}{3}\xi^2 +\ldots
                          , \quad \xi\ll 1\,, \nonumber\\
  &=& 2^{8/3}\Gamma(2/3)\xi^{-4/3} + \ldots ,   
 \qquad\xi\gg 1\,,
\ea
where $\xi=2\chi/3$ and 
$W_{\rm cl}=(2/3) \alpha m^2 \chi^2$ is the classical intensity
of synchrotron radiation. We glue the two approximations at $\xi=0.8$
together. The energy loss per time is $\beta=dE/dt=-W$.
We define as typical length-scale $\lsyn$ of synchrotron losses (in
the classical limit $\xi\ll 1$) 
\be \label{lsyn}
 \lsyn = \left( \frac{1}{E}\,\frac{dE}{dt}\right)^{-1}
 =\frac{3}{2\alpha m} \frac{1}{\gamma} \left( \frac{B_{\rm cr}}{B}\right)^2 \,.
\ee
\item
Below the pion-production threshold, $e^+e^-$ pair production on photons 
becomes for low magnetic field strengths the main energy loss mechanism 
of charged hadrons. We use the energy losses as calculated in
Ref.~\cite{Berezinsky:2002nc}. 
\item
For muons, we include energy losses due to inverse Compton scattering
from Ref.~\cite{inc}.
\end{enumerate}
In the following, we will consider static sources and neglect the
possible effects of adiabatic cooling as well as the energy losses due
to curvature radiation. The latter is formally equivalent to
synchrotron radiation after the replacement of the Larmor radius
$R_L=E/(eB)$ with the curvature radius $R_{\rm C}$ and needs therefore
no special consideration.

\subsection{Diffusion in magnetic fields}

The magnetic fields present in the surrounding of the acceleration
region induce not only synchrotron radiation as an important energy
loss process discussed in the previous subsection, but lead also to
deviations from straight-line propagation for charged particles.  The
transition between the ballistic and the diffusion regime happens
approximately at the energy $E_L$ when the Larmor radius,
\be
\label{eq:larmor}
 R_L = 1.08\times 10^{-3} {\rm pc} \;
       \frac{E}{10^{18}{\rm eV}} \: \frac{\rm G}{B} \,,
\ee
equals the  source size $R_s$, i.e.\ at the energy
\be
\label{eq:Ecr_sL}
E_L =  10^{18}{\rm eV} \: (R_s/{\rm pc}) \: (B/{\rm mG})\,.
\ee

In a turbulent magnetic field, the path of a charged particle can be
modeled either as a random-walk from a microscopic or as a diffusion
process from a macroscopic point of view. We shall use the latter
picture and need to specify thus the energy dependent diffusion
coefficient $D(E)$. Following Ref.~\cite{Aloisio:2004jd}, we use
\be
\label{D(E)}
 D(E) = D_0\left[ \left( \frac{R_L}{l_c}\right)^a
+  \left( \frac{R_L}{l_c}\right)^2 \right]\quad{\rm for}\;\; E<
 E_L \,, 
\ee
where $l_c$ denotes the coherence length of the magnetic field and
$a=1/3$ and $1$ correspond to the case of Kolmogorov and 
Bohm diffusion, respectively~\cite{D(E)}.

The diffusion time $\tdiff$ is defined as the typical time a
charged particle needs to diffuse the distance $R_s$ assuming
negligible energy losses,  
\be \label{tdiff}
 \tdiff=\frac{R_s^2}{6D} \,.
\ee
Thus the effective size $R_{\rm eff}$ of a magnetized source becomes 
$R_{\rm eff} = \tdiff = R_s^2/(6D)$. We choose the 
normalization constant $D_0$ in Eq.~(\ref{D(E)}) such that the effective
source size equals the true one, $R_{\rm eff}= R_s$, for $E_L$,
\be 
 D_0 = \frac{R_s}{6}\left(\frac{l_c}{R_s}\right)^2
 \frac{1}{1+(l_c/R_s)^{2-a}}\, .
\ee

Without diffusion, one defines the interaction depth as the
ratio $\tau_0=R_s/l_{\rm int}$ of the source size $R_s$ to the
interaction length $l_{\rm int}$ of nucleons with photons. For the
illustration of our numerical results, we determine in the following
$\tau_0$ via $l_{\rm int}=1/(n_\gamma\sigma)$ with $\sigma=0.2$~mb above
the threshold $E_{\rm th}$ as reference cross section. Therefore in
the case of a background of thermal photons with temperature $T$,
the interaction length becomes $l_{\rm int}\approx 2.5\times 10^{26}
\left(T/{\rm K}\right)^{-3}$~cm. In the case of diffusion, it is
convenient to introduce additionally an effective interaction depth
$\tau_{\rm eff}=R_{\rm eff}/l_{\rm int}$ or
\be
\label{eq:tau_eff}
 \tau_{\rm eff} = \left\{ \begin{array}{ll}
 \tau_0 & \quad {\rm for}\;\; E\geq E_L \\
 \tau_0 \left( \frac{E_L}{E} \right)^a
 \frac{1+(R_s/l_c)^{2-a}}{1+(R_L/l_c)^{2-a}} &
 \quad{\rm for}\;\; E<  E_L \,.
 \end{array} \right.
\ee
Thus a source that is transparent at high energies, when particles
move in the ballistic regime, becomes thick in the diffusion
regime below the energy $\sim E_L\tau_0^{1/a}$.

\subsection{Summary of interactions and propagation}

We idealize a neutrino source as an acceleration region surrounded by
a sphere of radius $R_s$ containing photons and turbulent magnetic fields.
The photon energy distribution may follow either a (broken) power-law
or a thermal Planck spectrum, although here we present results in
most cases only for a Planck  distribution.
The probability ${\cal N}$ that a particle diffuses outwards the
distance $\Delta r$ without scattering or decay is given by 
\be
 {\cal N} = \exp\left(-\int_{r}^{r+\Delta r} {\rm d}l\;
          (l_{\rm dec}^{-1} + l_{\rm int}^{-1}) \right)  \,,
\label{delta}
\ee
where $l_{\rm dec}$ and $l_{\rm int}$ are its decay and interaction length,
respectively\footnote{The corresponding equation in
  Ref.~\cite{I} contains a typo.}. The path length $\Delta l$ of the
trajectory and the distance $\Delta r$ diffused outwards are connected by
\be \label{nu}
 \nu(E) = \frac{\Delta l}{\Delta r} = \frac{R_s}{6D}  \,.
\ee
Thus diffusion enhances interactions and energy losses by the factor
$\nu(E)\propto 1/D(E)$ for $E\leq E_L$. The energy of the particle
along the path $l$ is obtained by integrating the energy losses $\beta(E)$. 
Note that in the calculation of diffusion coefficients one assumes
generally negligible energy losses. Thus, Eq.~(\ref{nu}) neglects the
decrease of the diffusion coefficient along the particle trajectory
and hence underestimates the interaction depth and the true number of
interactions.   

In our Monte Carlo simulation, we track explicitly all secondaries
($N,\pi^\pm,~K^\pm,~K^0_{L,S}$) for which the interaction rate is
non-negligible compared to their decay rate, and take into account the
possibility to produce prompt charm neutrinos; for details see~\cite{I}.


\section{Proton Flux}

For the illustration of our results we have chosen in the following
always for the initial proton flux a power-law  dependence, 
${\rm d}N_p/{\rm d}E = KE^{-\alpha_g}$ with $\alpha_g=2$, and
a rather high value of the maximal energy $E_{\rm max}=10^{24}$~eV.
In order to facilitate the comparison with our previous work Ref.~\cite{I}, 
we consider only thermal distributions of background photons.

Sources with negligible magnetic fields were thoroughly discussed in
Ref.~\cite{I}.  While in this case the initial and final proton fluxes
deviate strongly for thick sources, they basically coincide for
transparent ones. By contrast, the final proton spectrum is strongly
distorted even for a transparent source, if the magnetic field in
the source is sufficiently large:
At high energies, synchrotron radiation losses lead to a change in the
spectral index of the final proton flux, while at low enough energies
protons start to 
diffuse in the turbulent magnetic field. Thereby the number of interactions
increases,  the proton flux is suppressed and at the same time the
production of neutrons is enhanced. As we will show now, the cosmic ray flux
from a magnetized source has bumps and breaks, and may even suddenly
decrease by orders of magnitude.

\subsection{Low-energy protons and diffusion}
\label{subsec:Low-energy protons and diffusion}

\begin{figure}
\includegraphics[width=0.45\textwidth,angle=0,clip]{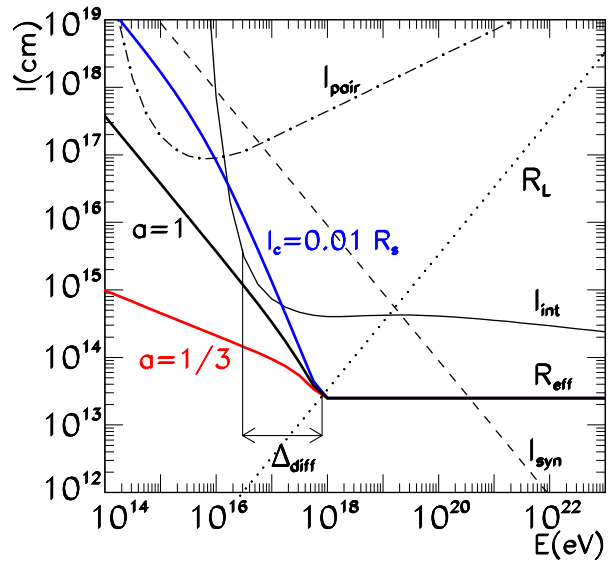} 
\caption{\label{fig:f_proton_t4_b1e2_tau.1_a1-.3_lc1e-2} Effective
  size (thick solid) of a source with a temperature of $T=10^4$~K,
  size $R_s=2.5\times 10^{13}$~cm ($\tau=0.1$), and magnetic field
  $B=100$~G for $a=1$ (black), $a=1/3$ (red), and with
  $l_c=10^{-2}R_s$ and $a=1$ (blue).  The interaction length (thin
  solid), the Larmor radius (dotted), and the typical length-scale of
  energy losses from synchrotron radiation (dashed) and $e^+ e^-$ pair
  production (dotted-dashed) in the case of protons are also shown.}
\end{figure}

We start our analysis by studying the consequences of diffusion
on the final proton flux. 
In Fig.~\ref{fig:f_proton_t4_b1e2_tau.1_a1-.3_lc1e-2}, we show the
effective size $R_{\rm eff}$ of a thin source with $T=10^4$~K,
$R_s=2.5\times 10^{13}$~cm ($\tau=0.1$) and a magnetic field strength
$B=100$~G for various different diffusion regimes: The case of
a magnetic field coherent over the entire source, $l_c=R_s$, is shown 
for Bohm diffusion, $a=1$, as a black solid and for Kolmogorov diffusion,
$a=1/3$, as a red solid line, respectively. 
Additionally, $R_{\rm eff}$ is shown for
$a=1$ and $l_c=10^{-2}R_s$ as a blue solid line and compared to other
typical length-scale as the interaction length, the Larmor radius
(dotted), and the synchrotron losses (dashed) for protons. The
behavior of  $R_{\rm eff}$ is the expected one: Protons with 
$E>E_L\approx 8\times 10^{17}$~eV propagate approximately on
straight-lines, and thus $R_{\rm eff}=R_s$. At lower energies, protons
start to diffuse and  $R_{\rm eff}$ increases.
The steepness of this increase depends on the energy dependence of the
diffusion coefficient $D(E)$, i.e.\ on $a$ and the value
of the coherence length $l_c$. For $R_L(E)\lesssim l_c$, the effective
size is proportional to $E^{-a}$. 
If the coherence length turns out to be smaller than $R_s$,
then $R_{\rm eff}$ is proportional to $E^{-2}$ in the energy range
from $E_L$ down to the energy where $R_L(E)=l_c$. In this case the
effect of diffusion is  strongest, see 
Fig.~\ref{fig:f_proton_t4_b1e2_tau.1_a1-.3_lc1e-2}.

\begin{figure}
\includegraphics[width=0.45\textwidth,angle=0,clip]{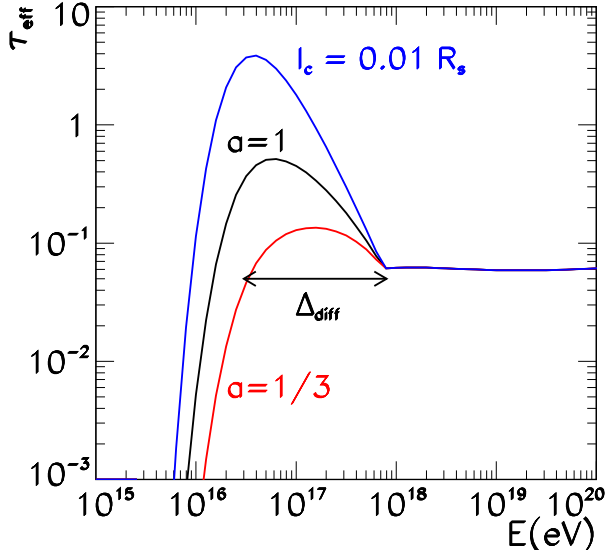} 
\caption{\label{fig:tau_eff_a1-.3_lc1e-2} Effective interaction depth
$\tau_{\rm eff}$ for the same source parameters as in
  Fig.~\ref{fig:f_proton_t4_b1e2_tau.1_a1-.3_lc1e-2}. 
}
\end{figure}

The increase of $R_{\rm eff}$ implies a change in the final proton flux,
only if $E_L$ is larger than the threshold energy $E_{\rm th}$ for
photo-pion production. In
Fig.~\ref{fig:f_proton_t4_b1e2_tau.1_a1-.3_lc1e-2},  we show the width
$\Delta_{\rm diff}\equiv \log(E_L)-\log(E_{\rm th})$ 
assuming $E_{\rm th}=m_p m_\pi/(2\varepsilon_\gamma)$, 
with $\varepsilon_\gamma=2.7T$. Inside this window, even a transparent
source can become opaque, as it is illustrated by the increase of the
effective interaction depth $\tau_{\rm eff}$ shown in
Fig.~\ref{fig:tau_eff_a1-.3_lc1e-2} in the case of diffusion.

\begin{figure}
\includegraphics[width=0.45\textwidth,angle=0,clip]{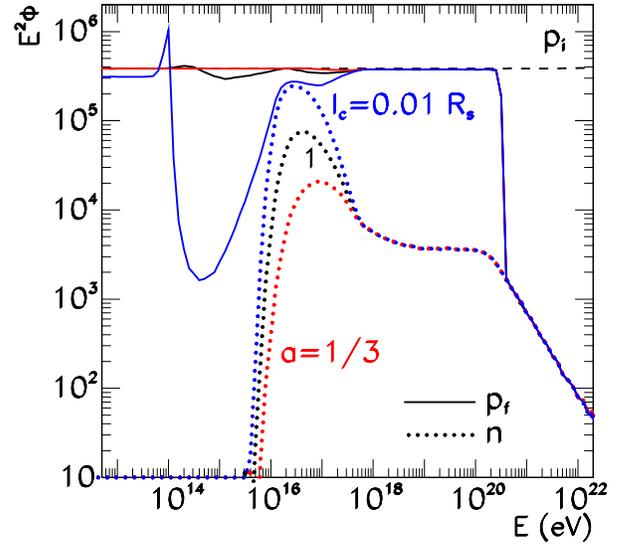}
\caption{\label{fig:fort.50_al2_t4_b1e2_tau.1_a1-.3_lc1e-2} 
Unnormalized fluxes of initial protons (dashed), neutrons escaping the
source before decaying (dotted), and the cosmic ray flux (sum of
escaped protons plus protons from neutron decays, solid line) for 
the same source parameters as in Fig.~\ref{fig:f_proton_t4_b1e2_tau.1_a1-.3_lc1e-2}.
}
\end{figure}

The impact of a change of $\tau_{\rm eff}$ on the final nucleon fluxes
is shown in Fig.~\ref{fig:fort.50_al2_t4_b1e2_tau.1_a1-.3_lc1e-2}, where
unnormalized fluxes of the initial and final protons as well as 
of neutrons escaping from the source before decaying are shown for the same
cases as above. 
The main effect of diffusion on the final proton spectrum is a
reduction of the flux in the energy window between $E_{\rm th}$ and
$E_L$. This reduction is proportional to the increase of the 
effective interaction depth as long as $\tau_{\rm eff}\lsim 1$.
Thus the dependence of the proton flux can be deduced directly from 
the behavior of $\tau_{\rm eff}$, which in turn depends on $R_{\rm eff}$.

In the case of neutrons, changes in the flux are also tightly
correlated to variations of $R_{\rm eff}$. The increase of $\tau_{\rm
eff}$ for protons implies more scatterings and therefore a larger
number of produced neutrons. Since the interaction depth of neutrons
is  not affected by diffusion, all neutrons escape from the source
before decaying, if the source is transparent. This
increased production of neutrons due to the diffusion  of protons
stops at $E_L$. At intermediate energies there is a flat plateau of
the neutron flux. Its slope is approximately the one of the initial
proton flux, only slightly modified by  the weak energy dependence of
the $p\gamma$ cross section.

\subsection{High-energy protons and energy losses}

At high energies, when the synchrotron loss length $l_{\rm syn}$
becomes smaller than the size of the source $R_s$, cf.
Fig.~\ref{fig:f_proton_t4_b1e2_tau.1_a1-.3_lc1e-2}, 
the proton flux is strongly suppressed. From Eq.~(\ref{lsyn}), this
limiting energy%
\footnote{The limiting energy $E_{\rm syn}$ should be distinguished from the
  maximal acceleration energy $E_{\max}$, defined by setting synchrotron energy
  losses equal to the energy gain in an electromagnetic field,
  $E_{\max}= [(3m^4)/(2\alpha^3 B)]^{1/2}$. The latter is evaluated
  for the magnetic field in the acceleration region.}
follows as
\begin{equation}
\label{eq:E_syn}
  E_{\rm syn} = \frac{3}{2\alpha} \left( \frac{B_{\rm cr}}{B}\right)^2 
  \frac{1}{R_s}\,.  
\end{equation}
Above this energy, the main component of the nucleon flux are neutrons
produced in scatterings. The energy losses of protons lead to a
steepening of the proton flux and therefore also of the neutron flux.
This steepening can be estimated for a transparent source by noting
that the energy of protons decreases with distance $l$ in the
classical limit as
\be
 y=\frac{E_f}{E_i} = \frac{1}{1+b_0 E_i l} \, ,
\ee
where $b_0 \equiv W_{\rm cl}/E^2 = (2/3)\alpha (B/B_{\rm cr})^2$.
Hence, the energy of a proton that interacts and produces a neutron is
reduced on average by the factor 
$\langle y\rangle =\int_0^R \!{\rm d} l\:y \propto 1/E_i$ (neglecting
logarithmic corrections). Therefore the exponent of the 
neutron spectrum is increased by one, $\alpha_n=\alpha_g+1$,
compared to the generation spectrum of protons.

Perhaps surprisingly, the final proton flux of a magnetized source is
not just a broken power-law, with $\alpha=\alpha_g$ below and
$\alpha=\alpha_g+1$ above $E_{\rm syn}$. Instead, there is
additionally a step in the flux at $E_{\rm syn}$, where, e.g., the
flux shown in Fig.~\ref{fig:fort.50_al2_t4_b1e2_tau.1_a1-.3_lc1e-2}
decreases suddenly by two orders of magnitude.  
The reason for this step is that the final proton flux at $E<E_{\rm syn}$
consists both of escaping ``direct'' protons and of the ones produced
in decays of escaping neutrons, whereas at $E>E_{\rm syn}$ only the
second contribution is significant. The difference in the relative
size of the final cosmic ray flux at  $E>E_{\rm syn}$ and $E<E_{\rm syn}$
corresponds to the ratio $R$ of the neutron and proton flux in the
plateau region, here at $10^{18}$~eV--$10^{20}$~eV.  It can be estimated for
transparent sources as $R_0\sim \tau_0\times{\rm BR}(p\to n)\times
(1-\langle y(\alpha_g)\rangle )$, where $1-\langle y(\alpha_g)\rangle$
is the spectrally averaged energy transfer from protons to neutrons.
If the interaction depth is large, protons produce neutrons before
losing too much energy due to synchrotron losses. Therefore, the step
at $E_{\rm syn}$, where the cosmic ray flux leaving the source changes
from neutron to proton dominated, increases for smaller $\tau_0$.
This step in the emitted cosmic ray flux at
$E_{\rm syn}$ acts in the case of rather transparent sources,
$\tau_0\ll 1$, effectively as the maximal energy of the source. The
distribution of maximal magnetic field strengths in different sources
and the resulting distribution of maximal energies ${\rm d} n/{\rm d}
E_{\max}$ leads thus to a steepening of the observed diffuse cosmic
ray flux and may thereby reconcile the observed steep spectrum
$\alpha_g\approx 2.6$ with the expectation $\alpha_g\approx$~2.0--2.2 from
Fermi shock acceleration~\cite{KS06}.

\subsection{Dependence on the magnetic field strength}
\label{subsec:Dependence on the magnetic field strength}
\begin{figure}
\includegraphics[width=0.45\textwidth,angle=0,clip]{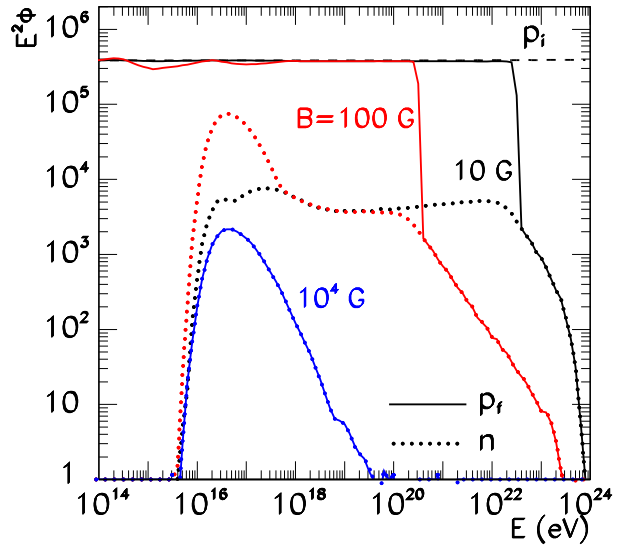}
\caption{\label{fig:fort.50_al2_t4_tau.1_b1e1-1e2-1e4_a1} 
Unnormalized fluxes of initial protons (dashed), neutrons escaping the
source before decaying (dotted), and the cosmic ray flux (sum of
escaped protons plus protons from neutron decays, solid line) for 
a source with a temperature of $T=10^4$~K, size $R_s=2.5\times
10^{13}$~cm ($\tau=0.1$), and magnetic field $B=10$~G (black),
$100$~G (red), and $10^4$~G (blue).}
\end{figure}
Let us discuss now the impact of the magnetic field strength on the
nucleon flux. Increasing $B$ has two main consequences: On the one hand,
the energy range $\Delta_{\rm diff}$ with diffusion increases, because
$E_L\propto B$. On the other hand, synchrotron radiation losses 
become important for lower energies, as $E_{\rm syn}\propto 1/B^2$. 
Thus an increase of the magnetic field strength widens the energy
range $\Delta_{\rm diff}$ with diffusion until
$E_L=E_{\rm syn}$. Increasing $B$ even further, $\Delta_{\rm diff}$
narrows again and synchrotron losses influence finally the whole
spectrum. This dependence is clearly illustrated by 
Fig.~\ref{fig:fort.50_al2_t4_tau.1_b1e1-1e2-1e4_a1}, where
the nucleon fluxes for a source with a  temperature of $T=10^4$~K,
size $R_s=2.5\times 10^{13}$~cm ($\tau=0.1$), and three different
magnetic fields $B=10$~G (black), $100$~G (red), and $10^4$~G (blue)
are shown. In the last case, the cosmic ray flux is strongly
suppressed and the main part of the source luminosity is damped into
electromagnetic cascades.

\subsection{Dependence on the interaction depth}
\label{subsec:Dependence on the thickness}
\begin{figure}
\includegraphics[width=0.45\textwidth,angle=0,clip]{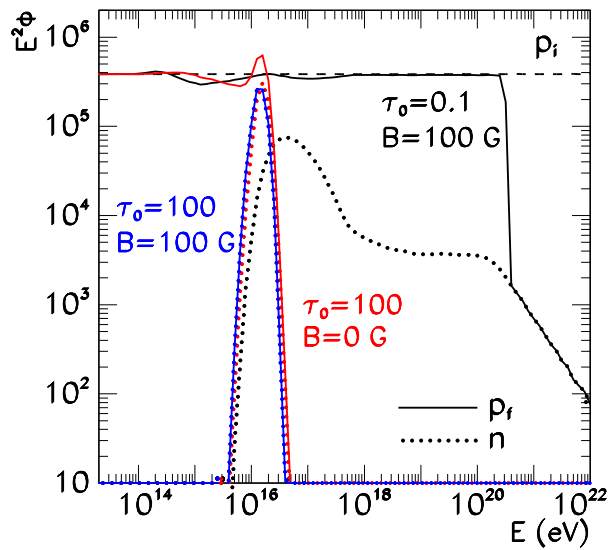}
\caption{\label{fig:fort.50_al2_t4_tau.1-1e2_b1e-5-1e2_a1.eps} 
Unnormalized fluxes of initial protons (dashed), neutrons escaping the
source before decaying (dotted), and the cosmic ray flux (sum of
escaped protons plus protons from neutron decays, solid line) for 
 a source with a
  temperature of $T=10^4$~K, size $R_s=2.5\times 10^{13}$~cm ($\tau=0.1$), and
  magnetic field $B=10^2$~G (black), $R_s=2.5\times 10^{16}$~cm
  ($\tau=10^2$) and $B=10^{-5}$~G (red), and $R_s=2.5\times 10^{16}$~cm
  ($\tau=10^2$) and $B=10^2$~G (blue).
 }
\end{figure}

Let us next consider the consequences of the magnetic field
strength on the nucleon flux from a thick source, $\tau_0\gsim 1$. 
As discussed in Ref.~\cite{I}, the characteristic property of a thick
source is the strong suppression of the nucleon flux at energies above
the photopion production threshold due to multiple scattering. For the
unmagnetized, thick source in
Fig.~\ref{fig:fort.50_al2_t4_tau.1-1e2_b1e-5-1e2_a1.eps}, 
the final proton flux is cut off at $E_{\rm th}\approx 10^{16}$~eV followed by a
small bump, while at lower energies the proton flux is hardly affected
as long as $R_s$ is smaller than the typical length scale for $e^+e^-$
pair production, see
Fig.~\ref{fig:f_proton_t4_b1e2_tau.1_a1-.3_lc1e-2}. 
The neutron flux is narrowly peaked around
$E_{\rm th}$, where the neutron interaction length decreases and the escape
probability increases.

The consequences of magnetic fields in a thick source are twofold. 
At high energies, there is a competition between multiple
scattering and  synchrotron radiation as the main energy loss
process. Analogous to Eq.~(\ref{eq:E_syn}), we define as
critical energy 
\begin{equation}
\label{eq:E_syn_thick}
  E_{\rm syn} = \frac{3}{2\alpha} \left( \frac{B_{\rm cr}}{B}\right)^2 
  \frac{1}{l_{\rm int}}\,.  
\end{equation}
While both processes lead to a strong suppression of the nucleon flux
at high energies, only the latter transfers the energy to neutrinos. 
In contrast,  magnetic fields influence not only the
neutrino but also the cosmic ray flux at low energies. Since
diffusion increases the effective size $R_{\rm eff}$ of the source
below $E_L$, synchrotron and $e^+e^-$ pair production losses can
suppress the proton flux even below the pion-production threshold. 
As result, the final cosmic ray flux may consist only of  
neutrons which escape near $E_{\rm th}$. As an example for such a
situation, we show the  proton and neutron flux  from a source with
$B=100$~G and $\tau_0=100$ in 
Fig.~\ref{fig:fort.50_al2_t4_tau.1-1e2_b1e-5-1e2_a1.eps} that is narrowly
concentrated around the threshold energy.

\section{Neutrino yields}

In order to analyze the effect of the magnetic fields of the source on
the neutrino spectrum we study now the neutrino yield of a single
source. The neutrino yield $Y_\nu(E)$, i.e. the ratio $Y_\nu(E) =
\phi_\nu(E)/(p_{\rm int}\phi_p(E))$ of the emitted neutrino flux $\phi_\nu$
and the product of the interaction probability $p_{\rm int}=1-\exp(-\tau_0)$
and the injected proton flux $\phi_p$, represents the number of
neutrinos produced per injected interacting proton with the same
energy\footnote{Note that this definition differs from the
  standard one, $Y_\nu(E)=\phi_\nu(E)/(\tau_0\phi_p(E))$, that was
  also used in Ref.~\cite{I}.}.

\subsection{General characteristics}
\begin{figure}
\includegraphics[width=0.45\textwidth,angle=0,clip]{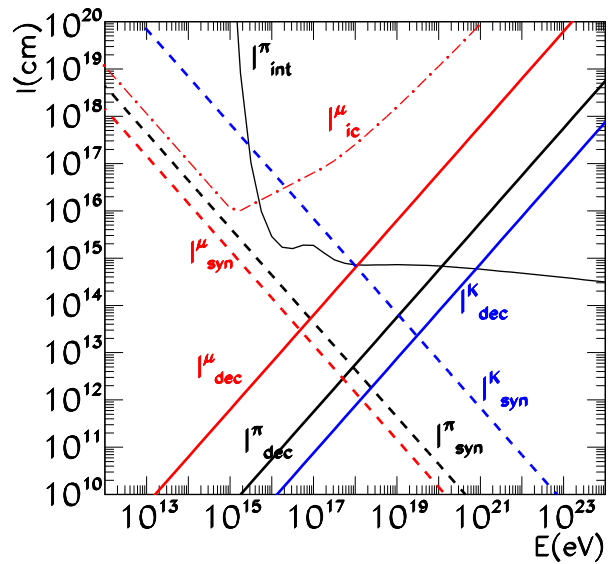}
\caption{\label{fig:f_nu_al2_t4_tau.1_b1e2_a1_p} Decay lengths (thick
solid) and the typical length-scale of synchrotron losses (dashed) for
charged pions (black), charged kaons (blue) and muons (red) for a
source with temperature $T=10^4$~K, size $R_s=2.5\times 10^{13}$~cm
($\tau=0.1$), and magnetic field $B=10^2$~G. The interaction length
for pions (thin solid) as well as the typical length of inverse
Compton scattering for muons (dotted-dashed) have been included.  }
\end{figure}
The dependence of the neutrino yields on the source parameters can be
deduced from the corresponding behavior of secondary mesons in the source.
Therefore we complement the typical length-scales shown in
Fig.~\ref{fig:f_proton_t4_b1e2_tau.1_a1-.3_lc1e-2}  for protons with
those for mesons in Fig.~\ref{fig:f_nu_al2_t4_tau.1_b1e2_a1_p}. We
have added the typical length of inverse Compton scattering for muons
as well as the decay lengths $l_{\rm dec}=\gamma l_{1/2}$ of mesons
and muons.

The most important difference with respect to a source with negligible
magnetic fields is the appearance of a length scale, 
$l_{\rm syn}\propto 1/E$, which decreases with energy (as long as
$\chi\ll 1$). Hence, the synchrotron length becomes for a magnetized
source at high enough energies always the smallest length scale, and
in turn meson decays and the neutrino flux will be suppressed.

We modify Eq.~(\ref{eq:E_syn}) to take into account possible
meson and muon decays, defining 
\begin{equation}
\label{eq:Esyn2}
E_{i,\rm syn} = \left\{ \begin{array}{ll}
\frac{3}{2\alpha} \left( \frac{B_{i,\rm cr}}{B}\right)^2 
  \frac{1}{R_s} &  {\rm for}\;\; f_i(B,R_s)\lesssim 1 \\ 
\left(\frac{3}{2\alpha} \left( \frac{B_{i,\rm cr}}{B}\right)^2 
  \frac{m_i}{l_{i,1/2}}  \right)^{1/2} &  {\rm for}\;\;
 f_i(B,R_s) \gtrsim 1  \, ,
\end{array} \right. 
\end{equation}
where the functions $f_i(B,R_s)$ are defined as
\begin{equation}
f_i(B,R_s) = \frac{2\alpha}{3} \left( \frac{B}{B_{i,\rm cr}}\right)^2
  \frac{R_s^2 m_i}{l_{i,1/2}} \, .
\end{equation}
The first expression represents the condition $\l_{\rm
  syn}(E_{\rm syn})=R_s$. At energies above $E_{\rm syn}$, charged
  mesons lose energy by synchrotron radiation. Mesons with energies
  between $R_s l_{1/2}/m$ and $E_{\rm syn}$  escape
  from the source before decaying. At energies smaller than $R_s
  l_{1/2}/m$ mesons decay inside the source.
The second expression corresponds to $\l_{\rm syn}(E_{\rm syn})=l_{\rm
  dec}(E_{\rm syn})$. In this situation mesons with energies below
  $E_{\rm syn}$ decay always within the source.

For given source parameters, the limiting energy $E_{\rm syn}$ above
which  synchrotron losses become crucial increases with mass and
decreases with the decay length of a particle. As a consequence, muons
and pions are affected already at much lower energies than kaons and
protons~\cite{Ando:2005xi,Asano:2006zz}.

\begin{figure}
\includegraphics[width=0.45\textwidth,angle=0,clip]{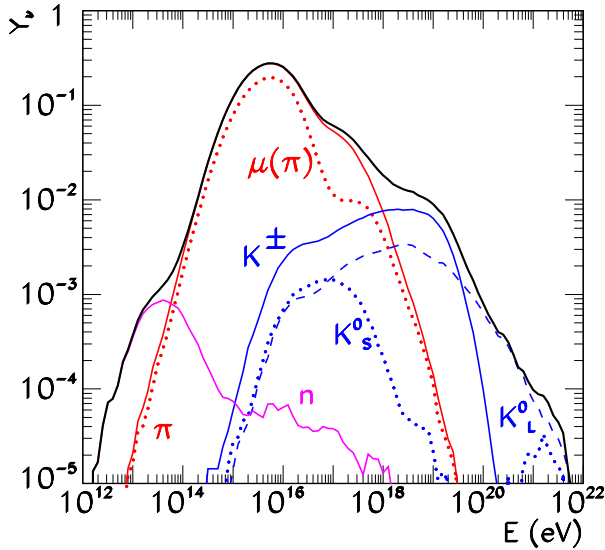}
\caption{\label{fig:nuyield_al2_t4_b1e2_tau.1_a1_p} Neutrino yield
  $Y_\nu$ from the decay of charged pions (solid red), charged kaons
  (solid blue), $K^0_L$ (dashed blue), $K^0_L$ (dotted blue), and
  neutrons (solid magenta). In the case pions we also show the
  fraction of neutrinos produced by the muons generated in the pion
  decay (dotted red).  The same source parameters as in
  Fig.~\ref{fig:f_nu_al2_t4_tau.1_b1e2_a1_p} have been assumed.  }
\end{figure}
In Fig.~\ref{fig:nuyield_al2_t4_b1e2_tau.1_a1_p}, we show how
different reactions contributing to the neutrino yield are affected by
magnetic fields. The increase of the number of $p\gamma$ interactions
at low energies by diffusion has two main consequences for the neutrino
production: First, a larger number of neutrons is produced, which will
freely escape in the case of a transparent source. Second, the number
of generated pions increases. The former leads to a bump of
$\bar\nu_e$ around the energy $\approx 10^{-3}E_{\rm th}$. The
latter produces another bump of neutrinos produced by pion decay
roughly at $E_{\rm th}$. These two characteristics can be clearly
observed in Fig.~\ref{fig:nuyield_al2_t4_b1e2_tau.1_a1_p}.

The high energy end of the spectrum is mainly influenced by
synchrotron losses. The relative importance of different channels to
the total neutrino yield is determined by the relative size of the
synchrotron energy, $E_{\rm syn}^\mu<E_{\rm
  syn}^\pi<E_{\rm syn}^K$, cf.~Fig.~\ref{fig:f_nu_al2_t4_tau.1_b1e2_a1_p}. 
Therefore, between the threshold energy for pion production and
$E_{\rm syn}^\pi$ pion decays are the most important contribution
to the neutrino yield. At slightly higher energies, between $E_{\rm
  syn}^\pi$ and $E_{\rm syn}^K$, neutrinos from the decay of charged
kaons dominate. For the source considered in
Fig.~\ref{fig:f_nu_al2_t4_tau.1_b1e2_a1_p},  all neutrinos are 
produced in the decay of $K^0_L$'s at the highest
energies, as they are not directly affected by
synchrotron losses. As discussed for neutrons, synchrotron losses of protons
lead however to a steepening of the $K^0_L$ spectrum and in turn also to
a steeper neutrino spectrum from these decays.
We  note that, in contrast to $K^0_L$, the $K^0_S$
contribution is suppressed roughly at $E_{\rm syn}^\pi$. Reason
for this is that the main decay channel of $K^0_S$'s giving rise to
neutrinos is $\pi^+\pi^-$.

Beside the neutron and pion neutrino bump at low energies and the
suppression at high energies due to synchrotron losses, there is a
third characteristic feature at intermediate energies: Muon damping
in the energy range between $E_{\rm syn}^\mu$ and $E_{\rm
  syn}^\pi$~\cite{Kashti:2005qa}.  
While the synchrotron length $l_{\rm syn}$ of pions and muons are
comparable, their decay lengths $l_{\rm dec}$ differ by
more than two orders of magnitude. As a consequence, muons with an
energy between  $E_{\rm syn}^\mu$ and $E_{\rm  syn}^\pi$ created from
pion decay can not further decay but lose energy by synchrotron
radiation. Hence, most  neutrinos in that energy range are $\nu_\mu$
produced  directly in pion decays.

In order to estimate the energy range where the effect of muon damping
is potentially strong, we define the muon damping width $\Delta_\mu$
analogous to $\Delta_{\rm diff}$ as
\begin{equation}   \label{delta_mu}
 \Delta_\mu \equiv \log(E_{\rm syn}^\pi) - \log(E_{\rm syn}^\mu) \, .
\end{equation}
The muon damping width depends on the source parameters $R_s$ and $B$
only indirectly via the choice between (\ref{eq:Esyn2}a) and
(\ref{eq:Esyn2}b). 
Let us first consider a source such that $f_\mu (B,R_s)\gtrsim 1$, i.e 
$(B/{\rm G})(R_s/{\rm cm})\gtrsim 3\times 10^{15}$. In this case, all
muons produced in pion decay with energies between $E_{\rm syn}^\mu$
and $E_{\rm  syn}^\pi$ are generated inside the source and therefore
are affected by the synchrotron losses. Hence, muons are completely
damped within this energy range. The muon damping width is
maximal and equals $\Delta_\mu \approx
0.5\times\log\left[(m_\pi/m_\mu)^5(l_\mu/l_\pi)\right]\approx
1.3$. If  $f_\mu (B,R_s)\lesssim 1$, some of the muons produced can
escape from the source before decaying.
Let us now consider a source such that $f_\pi (B,R_s)\lesssim 1$,
i.e. $(B/{\rm G})(R_s/{\rm cm})\ll 5\times 10^{14}$. In this case, 
pions with energies between $R_s l_{\pi,0}/m_\pi$ and $E^\pi_{\rm syn}$
decay outside the source, and then there is no muon damping.
The numerical values used in
Fig.~\ref{fig:nuyield_al2_t4_b1e2_tau.1_a1_p}, 
$(B/{\rm G})(R_s/{\rm cm}) = 2.5\times 10^{15}$, correspond to
partial muon damping, as can be observed in the figure.

A similar damping effect occurs for muons from charged kaon
decays. Now the numerical value of the width Eq.~(\ref{delta_mu}) is
larger. However, the effect is not as strong as in the case of pions,
because the energy range where kaons dominate is narrower and also
their number of decay channels is larger.

\subsection{Dependence on the properties of the source}

We now analyze how the general characteristics previously discussed
depend on the different properties of the source.

\subsubsection{Diffusion regime}

\begin{figure}
\includegraphics[width=0.45\textwidth,angle=0,clip]{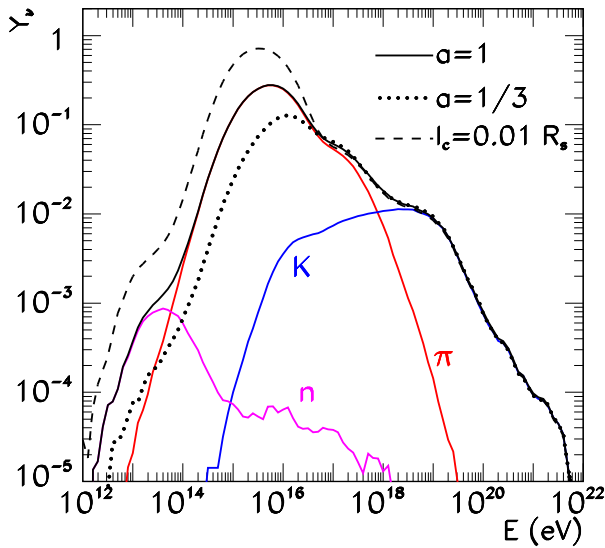}
\caption{\label{fig:nuyield_al2_t4_b1e2_tau.1_a1-.3-lc1e-2_pbis} Neutrino yield
  $Y_\nu$ produced in sources with the same parameters considered in
  Fig.~\ref{fig:f_proton_t4_b1e2_tau.1_a1-.3_lc1e-2}. 
The different
  reactions giving rise to $Y_\nu$ in the case of the source with Bohm
  diffusion ($a=1$) and   $l_c=R_s$ are also shown.
}
\end{figure}

In Fig.~\ref{fig:nuyield_al2_t4_b1e2_tau.1_a1-.3-lc1e-2_pbis}, we present
the neutrino yields for sources with different diffusion conditions,
choosing those already used in Sec.~\ref{subsec:Low-energy protons and
 diffusion}.
As previously discussed, diffusion generates two bumps, one from
neutrinos created by neutron decay at very low energies, and another
one of neutrinos produced in the decay of charged pions at $E\approx
E_{\rm th}$. The intensity of these peaks is directly related to
the characteristics of diffusion, in particular to $\tau_{\rm eff}$. 
A comparison of the yields with 
Figs.~\ref{fig:f_proton_t4_b1e2_tau.1_a1-.3_lc1e-2} 
and \ref{fig:tau_eff_a1-.3_lc1e-2} confirms the expected
correlation between the height of the neutrino bumps at low
energies and the effective interaction depth at the correct energy
range. Also, the different diffusion regimes do lead to the same
neutrino yield at high energies.

\subsubsection{Magnetic field strength}

\begin{figure}
\includegraphics[width=0.45\textwidth,angle=0,clip]{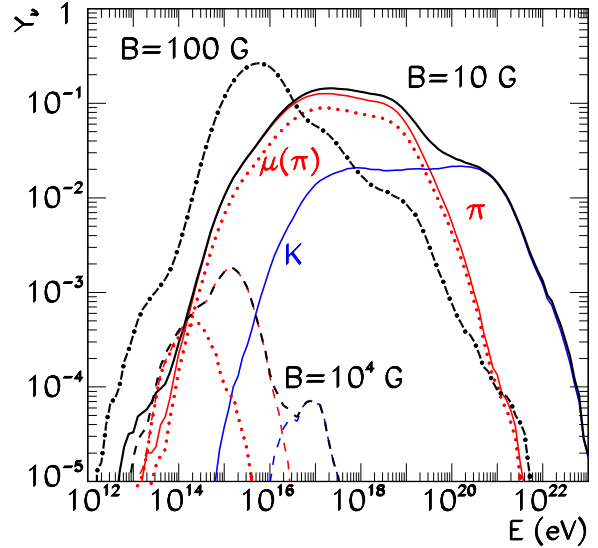}
\caption{\label{fig:nuyield_al2_t4_b1e1-1e2-1e4_tau.1_a1_pbis}
  Neutrino yield $Y_\nu$ produced in sources with the same parameters
  assumed in Fig.~\ref{fig:fort.50_al2_t4_tau.1_b1e1-1e2-1e4_a1}. The
  different contributions to $Y_\nu$ in the case of the source with
  $B=10$~G and $B=10^4$~G are also shown. The fraction of neutrinos
  produced by the muons generated in the pion decay (dotted red) is
  also indicated.  }
\end{figure}

In Fig.~\ref{fig:nuyield_al2_t4_b1e1-1e2-1e4_tau.1_a1_pbis}, we show
the dependence of the neutrino yield on the magnetic field strength. 
As discussed in Sec.~\ref{subsec:Dependence on the magnetic field
strength} two main effects are expected.  At low energies, the larger
the magnetic field strength the more intense is diffusion and hence
the interaction depth. This 
translates into an increase of the neutrino yield at low energies, and
in particular into a higher bump of the contribution to the neutrino
yield coming from pion and neutron decay. This can be easily confirmed
by comparing the two sources with 10~G and $100$~G in
Fig.~\ref{fig:nuyield_al2_t4_b1e1-1e2-1e4_tau.1_a1_pbis}.

The second consequence of a larger magnetic field strength is a more
severe suppression of the neutrino yield at high energies due to
synchrotron radiation. Since the energy at which synchrotron losses
become important is proportional to $1/B^2$, 
the suppression of the neutrino yield starts
earlier in the source with $10^2$~G than in the one with 10~G, while
for $B=10^4$~G  the neutrino yield is suppressed in the  whole
energy range,
cf. Fig.~\ref{fig:nuyield_al2_t4_b1e1-1e2-1e4_tau.1_a1_pbis}.

The value of the magnetic field strength also affects muon damping,
enhancing the damping for a stronger field. For the case of a source 
with 10~G shown in
Fig.~\ref{fig:nuyield_al2_t4_b1e1-1e2-1e4_tau.1_a1_pbis}, 
the product $(B/{\rm G})(R_s/{\rm
  cm})=2.5\times 10^{14}<5\times 10^{14}$. Therefore most
pions decay outside the source and thus muons are not affected
by the magnetic field. If the magnetic field is increased to $10^4$~G,
then $(B/{\rm G})(R_s/{\rm   cm})=2.5\times 10^{17}\gg 3\times 10^{15}$ 
and complete muon damping occurs.

\subsubsection{Interaction depth}

\begin{figure}
\includegraphics[width=0.45\textwidth,angle=0,clip]{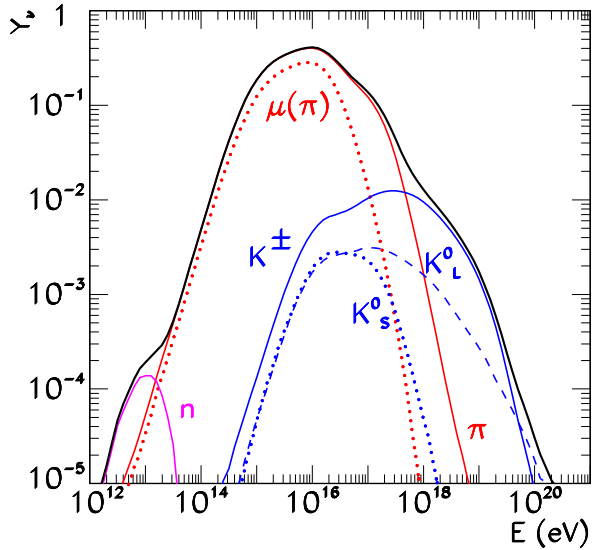}
\caption{\label{fig:nuyield_al2_t4_b1e2_tau1e2_a1_p} Neutrino yield
  $Y_\nu$ from the decay of charged pions (solid red), charged kaons
  (solid blue), $K^0_L$ (dashed blue), $K^0_L$ (dotted blue), neutrons
  (solid magenta), and charm mesons (solid green) for a source with
  temperature $T=10^4$~K, size $R_s=2.5\times 10^{16}$~cm
  ($\tau=10^2$), and magnetic field $B=10^2$~G. The fraction of
  neutrinos produced by muons from the pion decay (dotted
  red) is also indicated. } 
\end{figure}

We analyze now the influence of the interaction depth $\tau_0$ on a
magnetized neutrino
source. Figure~\ref{fig:nuyield_al2_t4_b1e2_tau1e2_a1_p} shows the 
different contributions to the neutrino yield for a source with
the same parameters as the one in
Fig.~\ref{fig:nuyield_al2_t4_b1e2_tau.1_a1_p}, but for 
$\tau_0=100$ instead of $\tau_0=0.1$.

At low energies, the narrowly peaked spectrum of neutrons leaving a
thick source is responsible for the absence of the high energy tail of
the neutrinos produced in neutron decay. The number of neutrinos from
pion decay in this range is not influenced strongly, because
multiple scattering occurs in both cases either because of
large  $\tau_{\rm eff}$ or large $\tau_0$.

At high energies there are some differences. In the case of thin
sources the most important contribution to the neutrino yield comes
from the decay of neutral kaons, see
Fig.~\ref{fig:nuyield_al2_t4_b1e2_tau.1_a1_p}, since they are not
directly affected by synchrotron losses. In the case of sufficiently
thick sources, also most neutral kaons interact with the photons
and thus the neutrino production from these decays is suppressed, see
Fig.~\ref{fig:fort.50_al2_t4_tau.1-1e2_b1e-5-1e2_a1.eps}.
In Ref.~\cite{I}, it was shown that neutrinos from decays of charmed
mesons became the dominant contribution at the highest energies for a
source with $\tau_0\gg 1$. In principle, the same holds in the case of
sources with magnetic fields. However,  synchrotron losses above
$E\gsim E_{\rm syn}$ may significantly suppress the proton flux and
thus also the production of charmed mesons produced. For the source
considered in
Fig.~\ref{fig:fort.50_al2_t4_tau.1-1e2_b1e-5-1e2_a1.eps}, the 
contribution from charm decays is negligible.

In the case of charged mesons and muons it is possible to redefine the
energy at which synchrotron radiation becomes the dominant energy loss
process substituting $R_s$ by $l_{\rm int}$ into  Eq.~(\ref{eq:Esyn2}).
If $f_\pi(B,l_{\rm int})\lesssim 1$ then pions with energies between
$l_{\rm int} m_\pi/l^\pi_0$ and $E^\pi_{\rm syn}$ scatter before
decaying. If $f_\pi(B,l_{\rm int})\gtrsim 1$ then pions decay inside the
source whereas muons with energies higher than $E^\mu_{\rm syn}$ will
suffer energy losses by synchrotron radiation before decaying. Therefore
the condition for having muon damping in a thick source can be written
as $f_\pi(B,l_{\rm int})\gtrsim 1$ or equivalently $(B/{\rm G})(l_{\rm int}/{\rm
  cm})\gtrsim 5\times 10^{14}$, which clearly satisfied by the source
considered in Fig.~\ref{fig:fort.50_al2_t4_tau.1-1e2_b1e-5-1e2_a1.eps}.

%

\section{Flavor dependence of neutrino yields}

Both neutrino telescopes and extensive air shower experiments have
some flavor discrimination possibilities. The case of neutrino
telescopes was discussed for the example of ICECUBE in detail in
Ref.~\cite{Beacom:2003nh}, for air-shower experiments see Ref.~\cite{EAS}.
The main observable for the neutrino flavor composition is
the ratio of track to shower events in a neutrino telescope, 
$R_\mu=\phi_\mu/(\phi_e+\phi_\tau)$, while only in a very small energy
range all flavors can be distinguished. Additionally, extensive air
shower experiments are sensitive to the fraction of tau events in all
horizontal neutrino events, $R_\tau=\phi_\tau/(\phi_e+\phi_\mu)$, in
a small energy window around $10^{18}$~eV.

In spite of these flavor discrimination possibilities, the potential
of high energy neutrino observations for mixing parameter studies
has been realized only recently~\cite{exc}, for some follow-up
studies see~\cite{exc2}. One of the main reasons for this missing
interest has been the prejudice that the maximal mu-tau mixing
together with the expected flavor ratio 
$\phi(\nu_e):\phi(\nu_\mu):\phi(\nu_\tau)=1:2:0$ from pion decay
prevents oscillation studies with high energy neutrinos.
Key observation of Refs.~\cite{exc} was that there exist however
several examples of neutrino sources where at least in some energy
range significant deviations from this canonical flavor ratio from
pion decay can be expected. Our results indicate that deviations 
from the canonical flavor ratio
$\phi(\nu_e):\phi(\nu_\mu):\phi(\nu_\tau)=1:2:0$ are much more common
than previously thought. In the following, we will
show that magnetized sources are characterized by a strongly
energy-dependent flavor ratio and thus the flavor ration encodes
non-trivial information.

\subsection{Energy dependence of the neutrino flavor ratio at the source}
\label{sec:flavorratio_source}

We start with an analysis of the expected neutrino flavor ratio
$R^0\equiv Y_{\nu_\mu}/Y_{\nu_e}$ at the source. 
\begin{figure}
\includegraphics[width=0.45\textwidth,angle=0,clip]{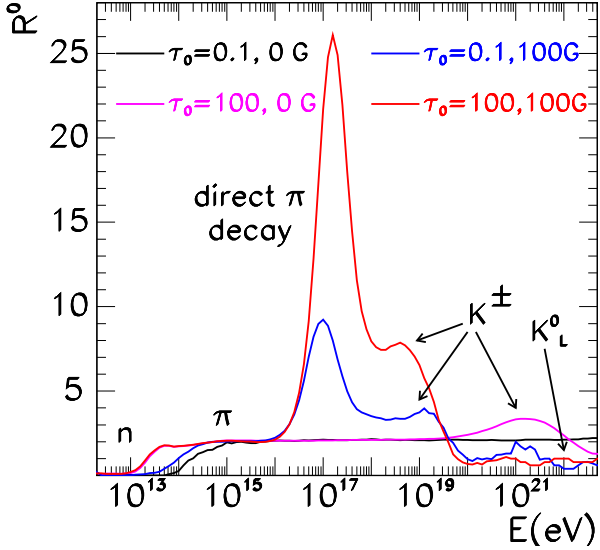}
\caption{\label{fig:flavorratio1} Flavor ratio $R^0$ of the neutrinos
  produced in a source at $T=10^4$, $B=0$~G, and $R_s=2.5\times
  10^{13}$~cm ($\tau   = 0.1$) (black), $B=0$~G, and $R_s=2.5\times
  10^{16}$~cm ($\tau   = 100$) (magenta),
$B=100$~G, and 
$R_s=2.5\times 10^{13}$~cm ($\tau
  = 0.1$) (blue), and $R_s=2.5\times 10^{16}$~cm ($\tau
  = 100$) (red) before they oscillate. The main
  reactions giving rise to the neutrino signal are also shown. The
  neutrinos produced in the muon damping are indicated as direct $\pi$
  decay. 
}
\end{figure}
In Fig.~\ref{fig:flavorratio1} we show the flavor ratio at the source
$R_0$ for four different sources with $T=10^4$: $B=0$~G, and
$R_s=2.5\times 10^{13}$~cm ($\tau = 0.1$) (black), $B=0$~G, and
$R_s=2.5\times 10^{16}$~cm ($\tau = 100$) (magenta), $B=100$~G, and
$R_s=2.5\times 10^{13}$~cm ($\tau = 0.1$) (blue),
$B=100$~G and $R_s=2.5\times 10^{16}$~cm ($\tau = 10^2$) (red).
In the case of a transparent source with negligible
magnetic field strength the flavor ratio  basically does not
depend on energy. However, the presence of magnetic fields as well
as large interaction depths can result in complicated structures of the
flavor ratio. The different bumps 
arising at different energies are tightly related to the different processes
taking place in the formation of the neutrino flux. 

Let us analyze these structures for each case. The flavor ratio $R_0$
of all sources is zero below the pion-production threshold as
most neutrinos are $\bar\nu_e$ produced in neutron decay. Right after
the energy threshold for pion production, $R_0$ becomes two, typical for
$\pi$ decay. 
At higher energies the behavior of the flavor ratio strongly depends
on the characteristics of the source.  Let us first summarize the two
cases with negligible magnetic fields.  If the source is transparent,
then $R_0$ remains constant, $R_0\approx 2$. Thick sources show, though, a
bump exactly at the energy at which there is a crossover between pion
and charged kaon neutrino dominance, namely, where $l^\pi_{\rm int}(E)
= l^\pi_{\rm dec}(E)$. At higher energies, the flavor ratio tends to
one, when all neutrinos come from the decay of charmed mesons, see
Ref.~\cite{I}.

In the case that sources have non-negligible magnetic fields the
energy dependence of the flavor ratio becomes more complex. The main
feature is the presence of a strong peak arising between $E^\mu_{\rm
  syn}$ and $E^\pi_{\rm syn}$, in this case around $E\approx
10^{17}$~eV. 
This clearly
reflects the energy window where muon damping occurs. In this case the
only neutrinos produced are the $\nu_\mu$ generated directly from pion
decay. The accompanying muons lose energy instead of decaying,
therefore no electron neutrinos are produced. As a consequence $R_0$
increases, the final value depending on the strength of the
muon damping. By comparing
Figs.~\ref{fig:nuyield_al2_t4_b1e2_tau.1_a1_p}
and~\ref{fig:nuyield_al2_t4_b1e2_tau1e2_a1_p}, one expects a stronger
effect for a thick source, as can be observed in the figure.
At higher energies there is an energy range where neutrinos are
produced in the decay of charged kaons, analogously to the case of
thick source with negligible magnetic fields. In contrast to this
case, though, the energy range of charged kaon dominance happens at
lower energies. The reason is that the crossover between pions and
charged kaon neutrinos takes place at $E^\pi_{\rm syn}$, which, for
the sources considered in Fig.~\ref{fig:flavorratio1}, is smaller than
the energy where $l^\pi_{\rm int}(E) = l^\pi_{\rm dec}(E)$.  The
different height of the bump observed in the transparent and thick
sources is related to the muon damping happening also for the charged
kaons in the latter case.
Finally at the highest energies one expects a tendency toward one,
either because these neutrinos are produced via $K^0_L\rightarrow \pi
+ e+ \nu_e$, as in the figure, or because they come from the decay of
charmed mesons.

We conclude that there is a direct relationship between the energy
dependence of the flavor ratio $R_0$ and the different processes
contributing to the neutrino yield, that may be used to deduce the intrinsic
characteristics of the source.

\subsection{Effect of neutrino oscillations}

The neutrino spectra at the source discussed in the preceding
subsection are modulated by oscillations. Therefore the expected
flavor ratios $R_i$ at the Earth are different from the original
ratios $R_i^0$ at the source. 

The neutrino fluxes arriving at the detector, $\phi^D_\alpha$, can be written
in terms of the initial fluxes $\phi_\alpha$ and the 
conversion probabilities $P_{\alpha\beta}$,
\begin{equation}
\phi^D_\alpha = \sum_\beta P_{\alpha\beta}\phi_\beta = P_{\alpha
  e}\phi_e + P_{\alpha \mu}\phi_\mu \,.
\end{equation}
Since the interference terms sensitive to the mass splittings $\Delta m^2$'s
do not contribute, the conversion probabilities are simply
\begin{equation}
P_{\alpha\beta} = \delta_{\alpha\beta}-2\sum_{j>k}\Re(U^\star_{\beta
  j}U_{\beta k}U_{\alpha j}U^\star_{\alpha k}) \,,
\end{equation}
where $U$ is the neutrino mixing matrix and Greek (Latin) letters are
used as flavor (mass) indices.

\begin{figure}
\includegraphics[width=0.45\textwidth,angle=0,clip]{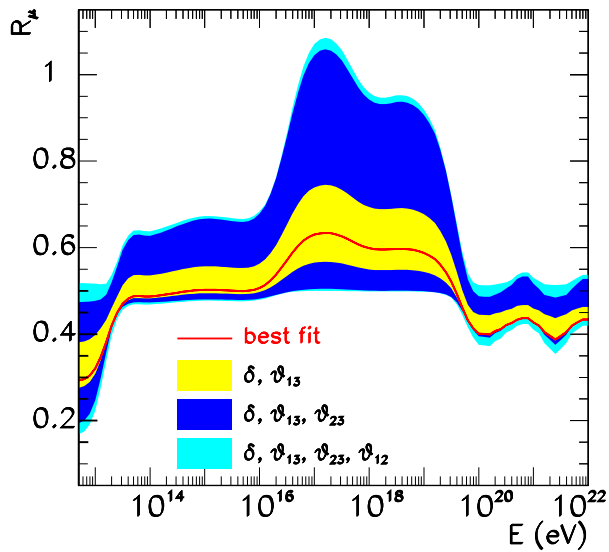}
\caption{\label{fig:flavorratio2} Flavor ratio $R_\mu$ at the Earth for
  a source with  $T=10^4$, $B=10^2$~G, and 
$R_s=2.5\times 10^{16}$~cm ($\tau
  = 10^2$). The solid red line (gray) corresponds to the best-fit point of the
neutrino mixing parameters, $\sin^2\theta_{12}=0.3$, $\sin^2\theta_{23}=0.5$ 
and $\theta_{13}=0$. In the colored areas mixing angles within 
95\%~C.L. and a possible non-zero value of $\delta_{\rm CP}$ have been
considered: $\theta_{13}$ and $\delta_{CP}$ in yellow (very light gray), plus
$\theta_{23}$ in blue (dark gray), and  $\theta_{12}$ in green
(light gray).
}
\end{figure}
In Fig.~\ref{fig:flavorratio2}, we show the expected flavor ratio $R_\mu$ 
at the
detector for a source with  $T=10^4$, $B=10^2$~G, and $R_s=2.5\times
10^{16}$~cm ($\tau  = 10^2$), as a
solid, red line for the best-fit point of the neutrino mixing
parameters, $\sin^2\theta_{12}=0.3,~\sin^2\theta_{23}=0.5$ and
$\theta_{13}=0$~\cite{nufit}. Various bands show the range of $R_\mu$
allowed if the mixing parameters are varied within their  95\%~C.L.,
while the unconstrained CP phase $\delta_{\rm CP}$
is  varied in the whole possible range, $\delta_{\rm CP}\in[0:\pi]$.
Dominant uncertainty for the prediction of $R_\mu$ is the
value of $\theta_{23}$. 

The main effect of neutrino oscillations is to smear out the structures
of the initial flavor ratio $R_0$. Nevertheless, the remaining energy
dependence still allows for an identification of the different regimes
described in Sec.~\ref{sec:flavorratio_source}. On the other hand, its
dependence on the neutrino mixing parameters offers the additional
possibility to obtain information on neutrino properties.

\section{Summary and conclusions}

We have calculated in this work the yield of high energy neutrinos produced in
astrophysical sources for arbitrary interaction depths $\tau_0$ and
magnetic field strengths $B$. Diffusion of protons increases their
path-length  and hence also the effective interaction
depth $\tau_{\rm eff}$. Therefore magnetized sources can lose
their transparency, while the neutrino fluxes can be increased in this
energy range. Sources with $\tau_{\rm eff}\gsim 1$ require to account
for multiple scatterings and for secondary meson-photon interactions.
For the latter task, we have extended the SOPHIA model in a
self-consistent way. 

Magnetic fields present in the surrounding of the acceleration region
lead not only to diffusion of charged particles, but induce also
synchrotron radiation as most important energy loss process.
Synchrotron losses result in a strong suppression of the cosmic ray
flux and hence also of the neutrino flux at high energies, $E\gsim
E_{\rm syn}$.

The cosmic ray upper bound of Refs.~\cite{WB,MPR} relies on, among
others, the assumption that the cosmic ray (without propagation
effects) and the neutrino flux have the same spectral shape.
The strong deviation of the escaping cosmic ray  and neutrino fluxes
from the injected power-law found by us therefore undermines the basis
of this bound even for transparent sources.

Since the relative importance of the various channels contributing to
the neutrino yields changes strongly as function of the energy, large
variations exist in the neutrino flavor composition emitted by a
magnetized source.  These variations are for magnetized sources even
stronger than those found previously for sources with negligible
magnetic fields in Ref.~\cite{I}.  In particular, we have examined two
specific examples discussed earlier in the literature, namely the
cases where the neutrino spectrum is dominated by kaon
decays~\cite{Ando:2005xi,Asano:2006zz} or influenced by the damping of
muons produced in pion decays~\cite{Kashti:2005qa}. In addition, we
have pointed out the possible existence of an analogous muon damping
in the case of charged kaon decay at high energies. We have analyzed
the conditions required for a source to present these features in the
neutrino yield. In particular, we have found that muon damping and
kaon dominance may influence the neutrino yields in the same source,
but in different energy ranges. 

The application of our simulation to concrete astrophysical models
and the calculation of diffuse neutrino fluxes will be performed in
paper III of this series~\cite{III}.

\section*{Acknowledgments}

SO and RT would like to thank NTNU, where part of this work was
done, for hospitality.  RT was supported by the Juan de la 
Cierva programme and by the Spanish grant FPA2005-01269.


\end{document}